# Cosmic Contact Censorship

Pushkar Ganesh Vaidya
Indian Astrobiology Research Centre (IARC), Mumbai, India.
pushkar@astrobiology.co.in

**Abstract**

Till date we have not made Contact with any Extraterrestrial civilizations. The Fermi Paradox remains a paradox - the Great Silence prevails. In the light of the Rare Earth hypothesis, it seems that advanced life is exceedingly rare in the universe. Also, if there does exist some Extraterrestrial civilizations then still achieving Contact is extremely difficult as the very nature of the universe discourages Contact. It's as if there is a Cosmic Contact Censorship in place.

**The Drake Equation**

Frank Drake developed the famous equation, named after him – the Drake Equation. It allows making calculated guesses at the number of advanced Extraterrestrial civilizations in our galaxy which are contactable through electromagnetic signals.

N = R* fp ne fl fi fc L **[1]**

where,

N = The number of communicative civilizations
R* = The rate of formation of suitable stars (stars such as our Sun)
fp = The fraction of those stars with planets. (current evidence indicates that planetary systems may be common for stars like the Sun.)
ne = The number of Earth-like worlds per planetary system
fl = The fraction of those Earth-like planets where life actually develops
fi = The fraction of life sites where intelligence develops
fc = The fraction of communicative planets (those on which electromagnetic communications technology develops)
L = The "lifetime" of communicating civilizations

When it was originally developed in the 1960's – the number of Extraterrestrial civilizations, N, was thought to be about 10,000. **[2]**

**The Fermi Paradox**

The Fermi Paradox (named after nuclear physicist Enrico Fermi who put forth the question in 1950) states that if there are Extraterrestrial civilizations, why haven't we met them? **[3]**

There are various solutions put forth for the Fermi Paradox. These solutions broadly fall into two categories;

1. Extraterrestrial civilizations do not exist (including they are obsolete already as well as they are yet to be born).

2. Extraterrestrial civilizations exist but there has been no interaction with us so far (due to variety of reasons including they have chosen not to do so in spite of being aware of our existence).

At the end of it all, so far we have not made Contact.

**Rare Earth Hypothesis**

The Rare Earth Hypothesis proposed by Peter Ward and Don Brownlee in their book *Rare Earth: Why Complex Life is Uncommon in the Universe* lists a variety of reasons to conclude that advanced life is extremely rare in the Universe. Broadly, it states: **[4]**

Microbial life is common in planetary systems
Advanced life (animals) is rare in the Universe

To quote from the Rare Earth homepage, 'The main conclusion of Rare Earth is that Earth is a very special place. Many circumstances and events had to happen, just right for Earth to remain a healthy habitat for advanced life. It appears our planet won the cosmic lottery* and we should cherish our very special place and time in the Universe'.

In short, it is very likely that we are alone!

Interestingly, the recent estimates of N for the Drake Equation are 1 or < 1, suggesting we are alone in the Milky Way. **[5]**

*The Rare Earth homepage actually carries a misprint stating 'comic lottery' instead of the intended 'cosmic lottery' (this misprint was carried in this paper until Vladimir Yershov, Mullard Space Science Laboratory pointed it out to me). I wonder which word is more appropriate!

**Cosmic Contact Censorship**

The status quo leads me to conceive an apparent Cosmic Contact Censorship.

In the light of the Rare Earth argument it appears we have only a handful of Extraterrestrial civilizations we can make Contact with in the universe, leave alone in our galaxy. Well, it's not easy to achieve Contact with them either.

The Weak Anthropic Principle (WAP) states that the physical constants and laws of the universe must be such as we measure them, i.e. these laws and constants must be compatible with our life. **[6]** These very factors also result in the universe being unsuitable for achieving Contact.

Here are the major limiting factors;

1. Our universe is large, very large (it's getting larger). The observable universe is at least 78 billion light years across. The large distances involved discourage intergalactic travel and communication. **[7]**

2. We cannot travel faster than the speed of light.

3. The strength of electromagnetic signals diminishes over large distances as they obey the inverse square law. In free space, doubling the distance from a signal source means the signal strength falls by almost a quarter. Also the signals are prone to disturbance due to various factors.

## Conclusions

It is evident; the very characteristics of the universe are not suitable for reasonable intergalactic and even interstellar, commuting and communicating.

The factors which make the universe suitable for life also make interaction between the possible different advanced life forms in the universe difficult.

It is as if there is an apparent Cosmic Contact Censorship in place.

## References

[1] http://www.setileague.org/general/drake.htm
[2] http://www.wired.com/wired/archive/12.12/life.html - The E.T. Equation, Recalculated - By Frank Drake
[3] A Solution to the Fermi Paradox: The Solar System, Part of a Galactic Hypercivilization? - Beatriz Gato-Rivera - arXiv:physics/0512062 v3 12 Jan 2006
[4] http://www.astro.washington.edu/rareearth/aboutthebook.html
[5] http://en.wikipedia.org/wiki/Drake_equation#Current_estimates_of_the_parameters
[6] On the Question of Validity of the Anthropic Principles - Zs. Hetesi, B. Balázs - arXiv:astro-ph/0609495 v1 18 Sep 2006
[7] Constraining the Topology of the Universe - Neil J. Cornish, David N. Spergel, Glenn D. Starkman, Eiichiro Komatsu - arXiv:astro-ph/0310233 v1 8 Oct 2003